\documentclass[10pt,prd,aps,twocolumn,a4paper,
floatfix,nofootinbib,showpacs,superscriptaddress]{revtex4-2}
\usepackage{ifpdf}
\ifpdf
\pdfoutput=1  %
\fi

\RequirePackage{snapshot}

\usepackage[utf8]{inputenc}
\usepackage{graphicx,psfrag}
\usepackage{graphics}
\usepackage[multidot]{grffile}

\usepackage{amssymb}
\usepackage{amsmath}
\usepackage{tabularx}
\usepackage{dcolumn}          %
\usepackage{physics}
\usepackage{mathrsfs}          %
\usepackage{amsmath,amsfonts,amssymb}
\usepackage{multirow}
\usepackage{comment}
\usepackage{hyperref} %
\hypersetup{
  colorlinks=true,
  linkcolor=blue,
  citecolor=blue,
  urlcolor=blue
}
\usepackage{bbm}
\usepackage{placeins}
\usepackage[normalem]{ulem}
\usepackage{tensor} %
\usepackage{verbatim}     %

\usepackage{xcolor}
\usepackage{orcidlink}

\def\Lie{\mathcal{L}}

\newcommand{\eg}{e.\,g.\@}
\newcommand{\ie}{i.\,e.\@}

\definecolor{cyan}{rgb}{0,0.9,0.9}
\definecolor{orange}{rgb}{0.9,0.5,0}
\definecolor{purple}{rgb}{0.8,0.4,0.8}
\definecolor{gray}{rgb}{0.8242,0.8242,0.8242}
\definecolor{grey}{rgb}{0.5,0.5,0.5}
\definecolor{pink}{rgb}{1.0, 0.0, 0.5}
\newcommand{\mytext}[1]{}

\makeatletter
\newcommand\ID[2]{#1\def\@currentlabel{#1}\label{#2}}
\makeatother
\renewcommand{\mytext}{}

\begin{document}

\title{Quasi-equilibrium configurations of binary systems of dark matter admixed neutron stars}

\newcommand{\affcentra}{Centro de Astrof\'{\i}sica e Gravita\c c\~ao -- CENTRA,
  Departamento de F\'{\i}sica, Instituto Superior T\'ecnico -- IST,
  Universidade de Lisboa -- UL, Av.\ Rovisco Pais 1, 1049-001 Lisboa,
  Portugal}
\newcommand{\affcoimbra}{CFisUC, Department of Physics, University of Coimbra, 3004-516 Coimbra, Portugal}
\newcommand{\affilfau}{Department of Physics, Florida Atlantic University, Boca Raton, FL 33431, USA}
\newcommand{\affilaei}{Max Planck Institute for Gravitational Physics (Albert Einstein Institute), Am Mühlenberg 1, Potsdam 14476, Germany}
\newcommand{\affilpots}{Institut für Physik und Astronomie, Universität Potsdam, Haus 28, Karl-Liebknecht-Str. 24/25, Potsdam, Germany}

\author{Hannes R. \surname{Rüter}\,\orcidlink{0000-0002-3442-5360}}
\affiliation{\affcentra}

\author{Violetta \surname{Sagun}\,\orcidlink{0000-0001-5854-1617}}
\affiliation{\affcoimbra}

\author{Wolfgang \surname{Tichy}\,\orcidlink{0000-0002-8707-754X}}
\affiliation{\affilfau}

\author{Tim \surname{Dietrich}\,\orcidlink{0000-0003-2374-307X}}
\affiliation{\affilpots}
\affiliation{\affilaei}

\date{\today}

\begin{abstract}

Using an adapted version of the \texttt{SGRID} code, we construct for the
first time consistent quasi-equilibrium configurations for
a binary system consisting of two neutron stars in which each is admixed with dark matter.
The stars are modelled as a system of two non-interacting fluids
minimally coupled to gravity. For the fluid representing
baryonic matter the SLy equation of state is used, whereas the second fluid,
which corresponds to dark matter, is described using the equation of state of
a degenerate Fermi gas.
We consider two different scenarios for the distribution of the dark matter.
In the first scenario the dark matter is confined to the core of the star,
whereas in the second scenario the dark matter extends beyond the surface
of the baryonic matter, forming a halo around the baryonic star.
The presence of dark matter alters the star's reaction to the companion's tidal
forces, which we investigate in terms of the coordinate deformation and mass
shedding parameters. The constructed quasi-equilibrium configurations
mark the first step towards consistent numerical-relativity simulations of
dark matter admixed neutron star binaries.

\end{abstract}

\maketitle

\section{Introduction}
\label{Section:Introduction}

In the present era of gravitational wave (GW) astronomy, the internal properties of compact stars can be probed during their mergers.
Using numerical-relativity (NR) simulations of the last stages
of a binary coalescence, it is possible to relate observational
GW data to properties of the source.
While these simulations have undergone significant improvements in the past, the impact of dark matter (DM) on the binary neutron star (NS) dynamics has not yet been investigated in detail and is not taken into account in standard GW analyses.
In fact, considering a coalescence of compact objects to occur in pure vacuum, could be an oversimplification that may lead to incorrect conclusions.

Due to their high compactness, NSs can trap and accumulate DM in their
interior throughout the star's evolution.
DM alters the compact star's properties, \eg{}, its mass, its radius,
its tidal deformability, its energy density and
speed of sound profiles~\citep{Goldman:1989nd,2006PhRvD..74f3003N,Ciarcelluti:2010ji,2011PhRvD..84j7301L,Leung:2012vea,Zurek:2013wia,Goldman:2013qla,Gresham:2017zqi,2018PhRvD..97l3007E,2019JCAP...07..012N,PhysRevD.102.063028,2020MNRAS.495.4893D,Das:2020ecp,Sagun:2021oml,DiGiovanni:2021ejn,RafieiKarkevandi:2021hcc,Leung:2022wcf,Rafiei_Karkevandi_2022,Sagun:2022ezx, Shakeri:2022dwg}.
Its effect depends on the relative fraction of DM and on the exact
equation of state (EoS) for the DM and baryonic matter (BM).
For an extended discussion of the impact of DM on compact star properties and its smoking gun signals, see Refs.~\citep{Maselli:2017vfi,Giangrandi:2022wht,Rutherford:2022xeb}.
While the effect of DM on isolated NSs can be probed through electromagnetic
observations, GW observations of binary systems of DM admixed compact stars
open up a new observational window and the possibility to probe a
density and temperature range larger that of isolated stars.
To push forward our understanding of the imprint of DM, we construct
quasi-equilibrium configurations of DM admixed NS binary system and
study the impact of DM focusing on quantities pertaining
to binary system, such as the orbital binding energy and the tidal deformations.

It is worth noting that not only NSs, but also black holes could be embedded into DM.
A step towards understanding the impact of DM on black hole mergers
was made in~\citep{Bamber:2022pbs}, where the behaviour of
wave DM around equal mass black hole binaries was studied in numerical
simulations.
Furthermore, GW signals from binary coalescences contain information of the binaries surrounding medium~\citep{Cole:2022fir}.

The effect of DM on the inspiral and post-merger phases of DM admixed NSs
has been studied by a few groups.
A first study by~\citet{Ellis:2017jgp} used a simple mechanical model,
and showed that a DM core can lead to the appearance of additional peaks
in the post-merger GW spectrum. In~\citep{Bezares:2019jcb} NR simulations of
equal-mass binaries consisting of BM admixed with a bosonic
Klein-Gordon field were performed.
For a DM mass fraction of 10\%, a redistribution of fermionic matter by the bosonic cores was found, followed by the formation of a
one-arm spiral instability. Another approach approximating
compact dark component as test particles was studied
in~\citep{Bauswein:2020kor}.
The simulations show the DM component to remain gravitationally bound after the merger of BM and orbit the center of the remnant with an orbital separation of a few km.
The DM core and a host star are likely to spin at different rotational
frequencies just after the merger due to the absence of non-gravitational
interaction. Further on, they may synchronise via the gravitational angular
momentum transfer, including tidal effects \citep{Hippert:2022snq}.

{The evolution equations for two-fluid binaries are quite well understood,
but so far no formalism for the construction of quasi-equilibrium
initial data exists.
Equations of motion for multi-fluid systems have been
derived in~\cite{AndCom21, And21} for the general case of interacting
fluids.}
Up to our knowledge, the first two-fluid NR simulations describing binaries
of DM admixed NSs were performed by~\citet{EmmSchPan22}
for a mixture of BM and mirror DM only interacting via the gravitational
field.
The results demonstrate that these systems tend to have a longer
inspiral phase with increasing amount of DM, which could be
associated to the lower deformability of DM admixed NSs.
These simulations however, did not start from initial data satisfying
the Hamiltonian and momentum constraints~\cite{Gou12,BauSha10,Tic17}
and the fluids did not start in
an equilibrium configuration. Instead the initial data was approximated
by superimposing Tolman-Oppenheimer-Volkoff (TOV)-like solutions~\citep{Tolman:1939jz,Oppenheimer:1939ne} of isolated DM admixed NSs.
In this work we construct consistent, constraint-solved,
quasi-equilibrium conditions for a two-fluid system of BM and DM.

One possible scenario for the formation of DM admixed NSs is the capture
of DM particles during the lifetime of the star, from a progenitor to the equilibrated NS stages.
The core of a NS is very dense and hence the chance of a DM particle
experiencing scattering is relatively high. In this scattering process
the particle transfers its kinetic energy to the star, becoming
gravitationally bound~\citep{Kouvaris:2013awa,Bell:2020jou,Bell:2020obw}.
This process is more efficient towards the Galactic center,
where the density of DM is many orders of magnitude greater than in the galaxy's
arms~\citep{2010PhRvD..82f3531K,2020Univ....6..222D,Nguyen:2022zwb}.
A conservative estimate of DM capture in the most central part of the Galaxy shows that stars accumulate up
to 0.01\% of DM during the main sequence and equilibrated NS stages
combined~\citep{PhysRevD.102.063028}.
However, also higher DM factions inside compact stars can be achieved through
other scenarios, e.g., DM production during a supernova explosion, accretion of DM clumps formed at the early stage of the Universe, or initial star formation on a pre-existing DM seed or local DM rich environments~\citep{Bertone:2007ae,Baryakhtar:2022hbu}.
If DM is symmetric, it cannot reach a high fraction due to self-annihilation, producing an electromagnetic or neutrino signal \citep{KouvarisThermalEvolution}. The latter scenario could lead to additional heating of isolated NSs as well as post-merger remnants~\citep{2010PhRvD..81l3521D,Hamaguchi:2019oev}, modification of kinematic properties \citep{2012PhLB..711....6P}.
Moreover, production of light DM particles, e.g., axions, in nucleon bremsstrahlung or in Cooper pair breaking and formation processes in the NS interior~\citep{PhysRevD.93.065044,Sedrakian:2018kdm,Dietrich:2019shr,Buschmann:2021juv}, could speed up the thermal evolution of a star by contributing an additional cooling channel.

We consider DM to be either concentrated in a core or extending beyond the
surface of BM, forming a DM halo around it.
As a first step, we consider non-interacting, fermonic DM
with spin $\frac{1}{2}$. The single star properties of this DM candidate have
been discussed in Ref.~\citep{PhysRevD.102.063028}.
The baryonic component is modelled through a piecewiese-polytropic
fit~\citep{Read:2008iy} of the SLy EoS~\citep{Douchin:2001sv} that reproduces nuclear matter ground state properties, fulfils heaviest pulsars measurements~\citep{PSRj03480432Article, Fonseca:2021wxt}, X-ray observations by
NICER~\citep{Miller_2019,Riley:2019yda,Raaijmakers:2019dks,Miller:2021qha,Riley:2021pdl},
and tidal deformability constraints from
GW170817~\citep{LIGOScientific:2018hze} and
GW190425~\citep{LIGOScientific:2020aai} binary NS mergers.

The two components interact only through gravity,
and therefore do not repel each
other, but overlap due to the absence of non-gravitational interaction.
This assumption is in very good agreement with the observations of the
Bullet Cluster~\citep{Clowe_2006, Randall:2008ppe} and direct DM
searches~\citep{Billard:2021uyg}, which show that the DM-BM cross section to be many orders
of magnitude lower than the typical nuclear one,
$\sigma_{DM-BM}\approx 10^{-45}\ \mathrm{cm}^2\ll \sigma_{BM}\sim10^{-24}\ \mathrm{cm}^2$.

By varying the particle mass and relative fraction of DM, we obtain either
a core configuration with a radius of the DM component less or equal to the
baryonic one, $R_{D}\leq R_{B}$,
or a halo with $R_{D} > R_{B}$~\citep{Sagun:2021oml}.
For both scenarios, we construct initial configurations employing
\texttt{SGRID}~\cite{Tic12,TicRasDie19}.
Many other codes exist for
the construction of quasi-equilibrium NS binary systems, notably
the spectral codes
\texttt{LORENE}~\cite{LORENE, GouGraTan01},
\texttt{Spells}~\cite{TacFouPfe15},
\texttt{FUKA}~\cite{FUKA,PapTooGra21},
\texttt{Elliptica}~\cite{RasFabBru22},
and the finite difference based code \texttt{COCAL}~\cite{UryTso12,TsoUryRez15,UryYosGou19}.
{In~\cite{PapTooGra21} the authors compared results from their independent
implementation with those 
from the SGRID code and find good agreement between both codes.}
Up to our knowledge, all codes mentioned above are only able to solve
systems consisting of a single fluid. Here we construct for the first 
time quasi-equilibrium binary configurations with two fluids.

The formalism and results are presented in geometric units in which
the gravitational constant $G=1$ and the speed of light $c=1$.
In these units, lengths are given as multiples of the solar mass,
$M_\odot$. For the conversion to SI units a spatial length must be multiplied
by $L_0$ = 1476.6250 $\mathrm{m}/M_\odot$ and
a time by $T_0$ = $4.9254909 \times 10^{-6}$ $\mathrm{s}/M_\odot$.
{In \autoref{tbl:geometric_units} we provide the conversion
to SI units for various quantities.}
Where appropriate we also use MeV to specify energy and mass of particles,
as well as SI units.
Throughout the paper,
Greek letter indices denote four dimensional, spacetime indices, whereas
Latin indices denote three-dimensional, spatial indices.

The paper is organized as follows.
In Section~\ref{Section:Formalism} we summarize the two-fluid formalism and DM distribution regimes.
Its implementation to the \texttt{SGRID} code is described in Section~\ref{Section:SGRID}.
In Section~\ref{Section:Results} we analyse the convergence properties of the
constructed configurations,
quantify the difference in the velocities of the two fluids
and investigate some physical properties of the quasi-equilibrium
configuration over a sequence of separations.
Section~\ref{Section:Conclusion} summarizes the results and discusses future
perspectives.

\begin{table}
  \caption{{Overview of the geometric units of various quantities 
  used in the text.}
  }
  \label{tbl:geometric_units}
  \begin{ruledtabular}
    \begin{tabular}{lll}    
      quantity & geometric units & SI units\\
      \hline
      length & $1 M_\odot$ & 1476.6250 $\mathrm{m}$ \\
      time   & $1 M_\odot$ & $4.9254909 \times 10^{-6}$ $\mathrm{s}$ \\
      velocity & 1 & 299792458 $\mathrm{m \, s^{-1}}$ \\
      mass   & $1 M_\odot$ & $1.98892 \times 10^{30}$ $\mathrm{kg}$ \\
      energy & $1 M_\odot$ & $1.78755 \times 10^{47}$ $\mathrm{J}$ \\
      specific enthalpy & 1 & $8.98755 \times 10^{16}$ $\mathrm{m^2 \, s^{-2}}$ \\
      angular momentum & $1 M_\odot^2$ & $8.80457 \times 10^{41}$ $\mathrm{kg \, m^2 \, s^{-1}}$ \\
    \end{tabular}
  \end{ruledtabular}
\end{table}

\section{Formalism}
\label{Section:Formalism}

We describe the matter as a system of two non-interacting perfect fluids
only indirectly coupled through the gravitational field.
This model is well justified, because the interaction between
standard model BM and DM is weak.
Utilisation of the perfect fluid model for DM is also justified,
as the mean free path and the scattering time scale of
DM particles can be small compared to the characteristic
time scales of the binary.
In the following, we estimate the mean free path and scattering time
in a semi-classical approach for a degenerate Fermi gas of particles.
{In this work we study a range of DM particle masses, but it is
only necessary to show the validity of the perfect fluid model
in the case farthest away from the hydrodynamical limit, \ie{}
for the most dilute DM component or equivalently for the largest 
mean free path.
For the configurations considered here, this is
configuration~\ref{id:2_dark_halo} in Table~\ref{tbl:configurations},
where the DM particle mass is}
170 MeV ($\approx 3 \times 10^{-28}\, \mathrm{kg}$).
The Fermi gas consists of non-interacting fermions, for which
a self-scattering cross section~$\sigma_{DM}$
formally does not exist.
Instead, we use the value of the upper limit
obtained from observations of merging galaxies, which yield
$\sigma_{DM}/m_p^{(DM)} < 1.25 \, \mathrm{cm^2/g} $,
with $m_p^{(DM)}$ the mass of the
DM~particles~\cite{Randall:2008ppe,Wittman:2017gxn}.
In this work we construct configurations
with a particle density~$n^{(DM)}$ of $0.7 \, \mathrm{fm^{-3}}$
in the center of the star.
Together with the upper limit for $\sigma_{DM}$ this
yields a mean free path $\lambda = 1/(n^{(DM)} \sigma_{DM})$
of $3.7 \times 10^{-17} \, \mathrm{m}$,
much smaller than the typical length scale of a NS,
which is on the order of $10^{4} \, \mathrm{m}$.
The scattering time scale can be estimated using the Fermi velocity,
which reaches values up to $0.8 \, c$ in the centre of the star.
Finally, using the value of the mean free path,
this yields a scattering time of
$t_c = \lambda / v_{DM} = 1.5 \times 10^{-25} \, \mathrm{s}$,
much smaller than for example the orbital period of the binary,
which in our configurations is a small as $3 \times 10^{-4} \, \mathrm{s}$.

At the surface of the stars DM reaches the free streaming limit
and the perfect fluid limit breaks down,
but there the density is so small,
that the impact on the gravitational field is low and hence the matter
in this region can be neglected.

For non-interacting fluids, the energy-momentum tensor can be split into
the two individual fluid components given by:
\begin{equation}
\label{eq:EnergyMomentumTensor1Fluid}
  T^{(s)}_{\mu\nu} = (e^{(s)} + p^{(s)}) u^{(s)}_\mu u^{(s)}_\nu
                     + p^{(s)} g_{\mu \nu} \, ,
\end{equation}
where $e$ is the proper energy density, $p$ is the pressure,
$u^\mu$ is the four velocity of the fluid and the label $(s)$ denotes the
particles species, which is either BM or DM.
The Einstein field equations are then given by
\begin{equation}
\label{eq:EinsteinFieldEquations}
  R_{\mu\nu} + \frac{1}{2} g_{\mu\nu} R
    = 8 \pi (T^{(BM)}_{\mu\nu} + T^{(DM)}_{\mu\nu} )
\end{equation}
and, because the two particle species do not interact, each fluid
satisfies the equations of motion of a single fluid. Consequently,
each fluid satisfies energy momentum conservation separately:
$\nabla^\mu T^{(s)}_{\mu\nu} = 0$.

For each fluid, we also define the rest mass density $\rho_0^{(s)}$,
which is computed from the number density $n^{(s)}$ by
\begin{equation}
  \rho_0^{(s)} = m_p^{(s)} n^{(s)} \, ,
\end{equation}
with $m_p^{(s)}$ being the mass of the particles.
Furthermore, the specific enthalpy is then given by
\begin{equation}
 h^{(s)} = \frac{e^{(s)} + p^{(s)}}{\rho_0^{(s)}} \, .
\end{equation}

To make the equations tractable, the
spacetime metric $g_{\mu\nu}$ is decomposed into a temporal
and a spatial part by introducing the spatial metric $\gamma_{ij}$,
the lapse $\alpha$, and the shift $\beta^i$~\cite{Coo00,BauSha10,Tic17}.
The line element in this 3+1 split reads
\begin{equation}
  ds^2 = -\alpha \,dt^2 + \gamma_{ij}\,(\beta^i dt + dx^i)(\beta^j dt + dx^j) \, .
\end{equation}
The extrinsic curvature $K_{ij}$ is related to the time derivative of
$\gamma_{ij}$, by the formula
\begin{equation}
  \label{eq:ExtrinsicCurvature}
  K_{ij} = - \frac{1}{2 \alpha}
    (\partial_t \gamma_{ij} - D_i  \beta_j - D_j  \beta_i) \, ,
\end{equation}
where $D_i$ denotes the covariant derivative compatible with the spatial
metric $\gamma_{ij}$.

We construct the partial differential equations governing quasi-equilibrium
by following the derivation in~\cite{Tic11}, which is trivially applied
to a system of non-interacting fluids.
To generate quasi-equilibrium configurations, we solve equations
for velocity potentials $\phi^{(s)}$, which are defined through the
following split of the four-velocity
\begin{equation}
\label{eq:VelocitySplit}
  \gamma^i_\mu u^{(s)\mu} =
   \frac{1}{h^{(s)}} ( D^i \phi^{(s)} + w^{{(s)} i} ) \, ,
\end{equation}
where $w^{{(s)} i}$ is a divergence free vector, i.e., $D_i w^{{(s)} i} = 0$,
describing the rotational part of the fluid.
Following the derivation of~\cite{Tic11}, we fix the time derivatives of
the fields by assuming the existence of
an approximate Killing vector $\xi$ and a set of
quasi-equilibrium conditions for the two fluids
\begin{align}
\label{eq:QuasiEquile}
  \Lie_\xi e^{(s)} \approx{}& 0 \, , \\
\label{eq:QuasiEquilp}
  \Lie_\xi p^{(s)} \approx{}& 0 \, , \\
\label{eq:QuasiEquilphi}
  \gamma^\mu_i \Lie_\xi (\nabla_\mu \phi^{(s)}) \approx{}& 0\, , \\
\label{eq:QuasiEquilw}
  \gamma^\mu_i \Lie_{\frac{\nabla \phi^{(s)}}{h^{(s)} u^{(s) 0}}} w^{(s)}_\mu
  \approx{}& 0 \, .
\end{align}
We omit further details of the derivation, since for non-interacting
fluids everything can be directly carried over to the individual fluid
components, and we state only the resulting
partial differential equation for the velocity potentials $\phi^{(s)}$:
\begin{equation}
\label{eq:PDEContinuityEquation}
  D_i \left ( \frac{\rho^{(s)}_0 \alpha}{h^{(s)}}
               (D^i \phi^{(s)}  + w^{(s) i} )
             - \rho^{(s)}_0 \alpha u^{(s) 0} (\beta^i + \xi^i)
      \right ) = 0 \, ,
\end{equation}
where the boost factor $u^{(s) 0}$ is given by
\begin{equation}
\label{eq:BoostFactor}
  u^{(s) 0} = \frac{\sqrt{{h^{(s)}}^2
  + (D_i \phi^{(s)}  + w^{(s)}_i )
    (D^i \phi^{(s)}  + w^{(s) i} ) }}{\alpha h^{(s)}} \, ,
\end{equation}
and the specific enthalpy is given by the expression
\begin{equation}
\label{eq:SpecificEnthalpy2}
  h^{(s)} = \sqrt{{L^{(s)}}^2 - (D_i \phi^{(s)}  + w^{(s)}_i )
             (D^i \phi^{(s)}  + w^{(s) i} )} \, ,
\end{equation}
with
\begin{equation}
\label{eq:LSquared}
  {L^{(s)}}^2 =
    \frac{b^{(s)} + \sqrt{{b^{(s)}}^2
        - 4 \alpha^4 ((D_i \phi^{(s)}  + w^{(s)}_i ) w^{(s) i})^2}}{2 \alpha^2} \,
\end{equation}
and
\begin{equation}
\label{eq:b}
  b^{(s)} = ((\xi^i + \beta^i) D_i \phi^{(s)} - C^{(s)})^2
             + 2 \alpha^2 (D_i \phi^{(s)}  + w^{(s)}_i ) w^{(s) i} \, .
\end{equation}
The variable $C^{(s)}$ is a constant,
which can vary for each star and which controls the mass of the fluid component.

For the approximate Killing vector $\xi^i$ we make the following ansatz:
\begin{equation}
\label{eq:KillingVector}
  \xi^i = \Omega ( -y, x - x_{CM}, 0) + \frac{v_r}{D} (r^i - r^i_{CM}) \, ,
\end{equation}
where $\Omega$ is the instantaneous orbital frequency,
$D$ is the separation between the star centres,
$v_r$ is the radial velocity, and $x_{CM}$ is the
$x$-coordinate of the centre of mass.

At apsis the orbital frequency together with the separation of the stars
control the orbital parameters like eccentricity and length of
the semi-major axis. Away from apsis there is a non-vanishing
radial component of the velocity to be taken into account.
In cases like the ``circular'' inspiral there is no apsis, but there
{ is a small, but non-vanishing, radially inward directed velocity component $v_r$.
There exist analytic approximations from Effective-One-Body- or Post-Newtonian theory, which provide a way to obtain low-eccentricity 
configurations~\cite{KiuSekShi09}. However, 
those expressions are derived in coordinates, that are not trivially 
related to the coordinates used in the extended
conformal thin sandwich (XCTS) formalism~\cite{BauSha10,Tic17}. 
Hence, to obtain really ``circular'' inspirals, in
practice $v_r$ must be obtained through eccentricity reduction~\cite{KyuShiTan14} evolutions of the data and adjusting $\Omega$ and 
$v_r$ appropriately.}

The configurations presented in this work are constructed within the
quasi-circular approximation for which the radial component is neglected,
$v_r=0$.
{ This approximation is well justified, because the change in 
orbital separation $\Delta D$ during one orbit is much smaller than the 
orbital period $T$.
Even a few orbits before merger $\Delta D$ is typically more than 100 
times smaller than $T$.}

We set the value of $\Omega$ to its value at
second Post-Newtonian order in
Arnowitt-Deser-Misner (ADM) gauge~\cite{SchWex93,TicBruCam02,TicBru03}.
{ $\Omega$ is then a function of the stellar masses and the orbital
separation $D$.
For the stellar masses we use} the sum of the
rest masses of the two fluids, which are computed by
\begin{equation}
  m_{0i}^{(s)} = \int_{V_i} \rho_i^{(s)} u^{(s) 0} \alpha
                 \sqrt{\det(\gamma_{jk})} d^3 x  \, ,
\end{equation}
where $V_i$ is the spatial volume over which the $i$-th star extends.
The value of $x_{CM}$ is then given by
\begin{equation}
  \label{eq:xCM}
  x_{CM} = \frac{(m_{01}^{(BM)}+m_{01}^{(DM)}) x_{c1}
               + (m_{02}^{(BM)}+m_{02}^{(DM)}) x_{c2}}
               {m_{01}^{(BM)}+m_{01}^{(DM)}+m_{02}^{(BM)}+m_{02}^{(DM)}} \, ,
\end{equation}
where $x_{c1/2}$ are the $x$-coordinates of the centres of the stars.
In this work, we present results for
equal-mass configurations only, \ie{}, $x_{CM} = 0$.

Besides the continuity equation (Eq.~\eqref{eq:PDEContinuityEquation})
governing the
fluid velocity potentials $\phi^{(s)}$, the metric must be fixed in a way
satisfying the ADM constraints.
To this end we choose a conformally flat ansatz for the spatial metric,
\ie{}, $\gamma_{ij} = \psi^4 \bar \gamma_{ij}$,
with $\gamma_{ij} = \delta_{ij}$ and $\partial_t \gamma_{ij} = 0$,
and construct the data on maximally sliced hypersurfaces,
\ie{}, the trace of the extrinsic curvature vanishes:
 $K=0$ and $\partial_t K$ = 0.
The free metric components are the lapse, shift, and conformal factor $\psi$
and their governing equations are formulated in terms of the XCTS equations~\cite{BauSha10,Tic17}.
Together with Eq.~\eqref{eq:PDEContinuityEquation}, the data is constrained
by a set of seven coupled partial differential equations, which
are solved iteratively one-by-one in a self-consistent manner.

\section{SGRID}
\label{Section:SGRID}

We have adapted the pseudo-spectral
\texttt{SGRID} code~\cite{Tic12,TicRasDie19}
to generate quasi-equilibrium configurations for two fluid systems.
We use the same iteration scheme that is used in~\cite{TicRasDie19}
for single-fluid NSs.
We sketch the iteration scheme in the following
with an emphasis on the adaptions and changes made.

\begin{enumerate}

\item
To ensure the convergence of the solver, it is necessary to provide
an initial guess sufficiently close to the true solution.
This initial guess is chosen as a superposition
of two boosted TOV-like two fluid stars of a given mass.
To generate solutions with particular rest masses for the
fluid components, one has to find the central pressures for which the
masses are realized. Since we are dealing with two fluids, this is
a two-dimensional root finding problem. In our tests, we found that using the
Newton-Raphson method is not always reliable, because the masses are not
a monotonous function of the central pressures, hence, a Newton-Raphson solver
easily gets caught in a local extremum of the mass function.
Instead, we employ a series of bisections on the central pressure of
one fluid component while keeping the central pressure of
the other fluid fixed.
The series of bisections iterates between the two fluid components
in a self-consistent manner until the fluid masses are sufficiently
close to the target parameters.

\item
\label{item:IterationSolvePhi}
If the residuals of Eq.~\eqref{eq:PDEContinuityEquation}
are larger than 10\% of the combined residuals of the XCTS equations,
we solve Eq.~\eqref{eq:PDEContinuityEquation}
and set the new $\phi^{(s)}$ to be the average of the old solution
$\phi^{(s)}_\mathrm{old}$ and the just obtained solution
$\phi^{(s)}_\mathrm{ell}$, using the following weights
$\phi^{(s)} = 0.8 \phi^{(s)}_\mathrm{old} + 0.2 \phi^{(s)}_\mathrm{ell}$.

\item
We proceed by solving the XCTS equations and update $\alpha$, $\beta$,
and $\psi$ in the same way, averaging the old and new solution.

\item
We do not adjust the values of $\Omega$ and $x_\mathrm{CM}$ as
in~\cite{TicRasDie19}. The value of $\Omega$ would be fixed within
an eccentricity reduction scheme. $x_\mathrm{CM}$ is left at its
Newtonian value, Eq.~\eqref{eq:xCM}.

\item
\label{item:IterationUpdateh}
We adjust the four constants $C^{(s)}$,
such that the rest masses of each component and in each star
match our desired target masses.
We then update the values of $h^{(s)}$ keeping it fixed until the end
of the next iteration.

\item
If the sum of the
residuals is below a certain tolerance or a prescribed maximum number
of iterations is reached, the iteration ends here and
is concluded with a final solving of the XCTS equations.

\item
The system of partial differential equations does not fix the
position of the stars and, hence, they will slowly
drift if not kept under control.
To keep the stars in place, the center of the stars are driven
back to the desired position.
For single fluids, the center is usually defined in an unambiguous way
as the point of maximum density. For two fluids the definition is
ambiguous, because the tidal deformations due to the companion star
are different for each fluid component and, consequently, the maximum
densities are at different points. In most cases, however,
the two maximum points will still be close.
The results shown in this work are obtained by choosing the point
with the maximum of the total proper energy density,
$e^{(tot)} = e^{(BM)} + e^{(DM)}$,
as the center of the stars.
We have chosen $e^{(tot)}$, in particular,
because it is a covariant scalar and it is the major quantity
determining the gravitational potential, hence giving an estimate
for the center of mass of the star.
To drive the center of mass back, the values of $h^{(s)}$ are transformed
by
\begin{equation}
  h^{(s),\mathrm{new}} = h^{(s)} + \Delta r^i \partial_i h^{(s)} \, ,
\end{equation}
where $\Delta r^i = r^i_\mathrm{current} - r^i_\mathrm{desired}$.

\item
Continue with step~\ref{item:IterationSolvePhi}.

\end{enumerate}

\begin{figure}
  \includegraphics[width=\columnwidth]
  {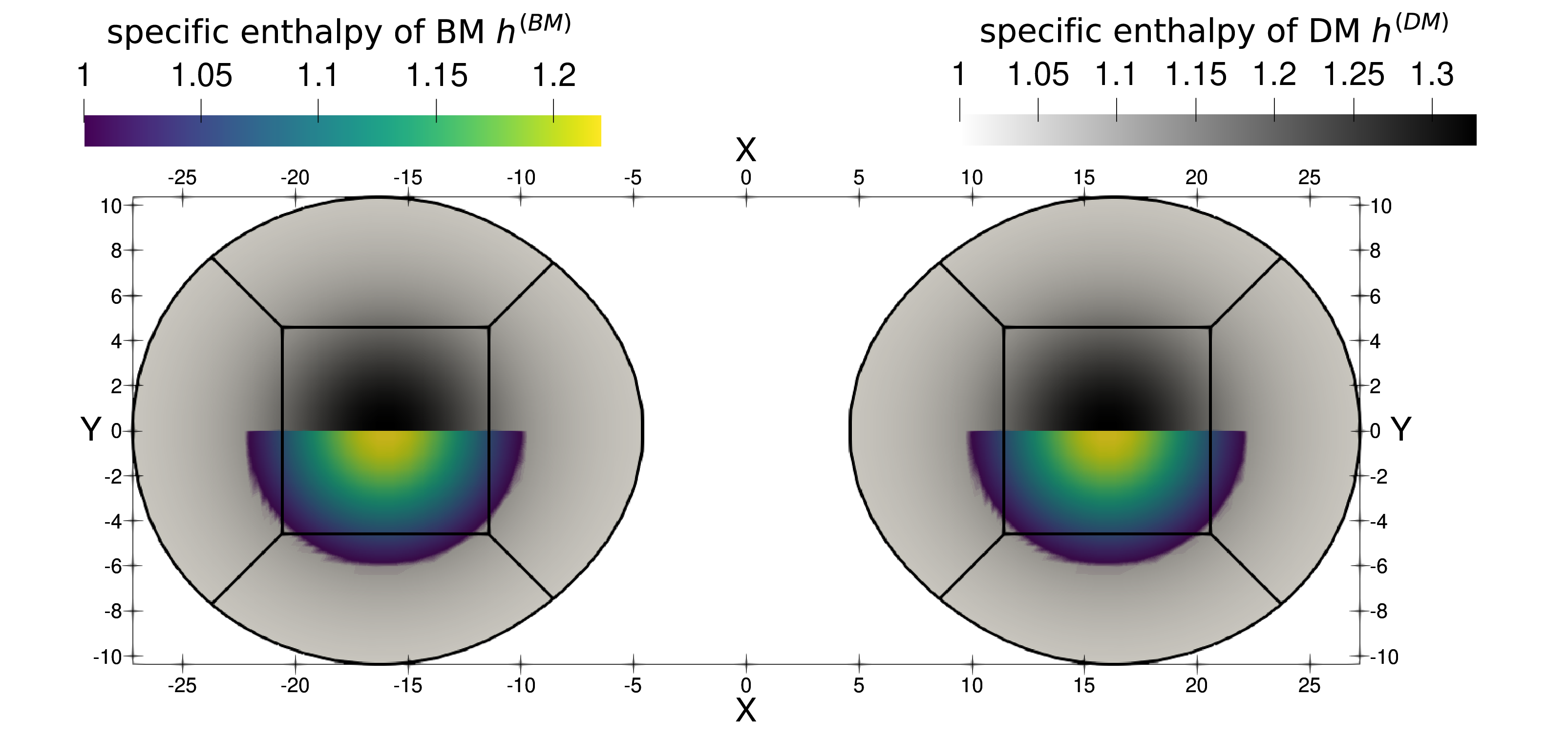}
  \caption{
    \label{fig:SpecificEnthalpy2D} Specific enthalpy in the
    $z=0$ plane for a configuration with DM halo.
    In the upper halves only the specific enthalpy of DM is shown,
    whereas in the lower halves the BM component lies on top of it.
    The black lines indicate the boundaries of the
    spectral elements. Each NS is comprised of a central cubical element
    and six cubed sphere elements (of which only four intersect the
    $z=0$ plane).
    {The DM particle mass in this configuration is 170 MeV 
    (corresponding to ID~\ref{id:2_dark_halo} in Table~\ref{tbl:configurations})
    and the} separation between the NS centres 
    amounts to 32~$M_\odot$ (47.3~km).
    }
\end{figure}

The \texttt{SGRID} code uses surface-fitted coordinates to reduce the Runge phenomenon
at the surface of the star. Each time we update the specific enthalpy
$h^{(s)}$ (step~\ref{item:IterationUpdateh} in the iteration),
we adapt the grid such that the boundaries of spectral elements
coincide with the new surface of the outer fluid.
That means we only construct configurations in which the
surfaces of the two fluids do not intersect, which would in principle
be possible given the different deformabilities of the fluids.
Furthermore, we do not construct domains that are adapted to
the surface of the inner fluid. Therefore, at the surface of the
inner fluid one can expect to observe the Runge phenomenon and a
slight degradation of the convergence in the truncation error.
Fig.~\ref{fig:SpecificEnthalpy2D} shows a visualisation of the
deformed spectral elements inside the NS and the distribution of
matter in terms of the specific enthalpy, {for a configuration with
a DM particle mass of 170 MeV.}

To close the system, the EoS is required to relate
$e^{(s)}$, $p^{(s)}$, $\rho_0^{(s)}$, and $h^{(s)}$.
For the EoS, \texttt{SGRID} reads in either parameters of
piecewise polytropes or EoS tables.
EoS tables are interpolated in a thermodynamically consistent
manner~\cite{Swe96} using a cubic Hermite interpolation.
To find the thermodynamic quantities for a given specific enthalpy
a Newton-Raphson root finder is used.
At low densities we use a polytrope that is matched at the lowest density
of the table.

{We validated our implementation of the TOV equations and 
the EoS interpolation by comparison of the SGRID implementation and the code used in~\cite{PhysRevD.102.063028}. 
We find that the TOV-like solutions of the two implementations deviate
only by machine round-off.}

\section{Results}
\label{Section:Results}
\subsection{Parameters of Constructed Configurations}
\label{Subsection:Configurations}

\begin{table*}
  \caption{{Properties of the used isolated NS. 
  All configurations have the same total ress mass: $m_0^{(BM)} + m_0^{(DM)} = 1.4 M_\odot$.
  The ID is a number used for reference in the text. 
  $m_p^{(DM)}$ is the DM particle mass and $m_0^{(DM)} / (m_0^{(BM)} + m_0^{(DM)})$ is the rest mass fraction of DM.
  $m^{(BM)}$ and $m^{(DM)}$ are the gravitational masses of the BM and DM component respectively.
  $R^{(BM)}$ and $R^{(DM)}$ are the radii of the BM and DM surface in Schwarzschild coordinates.}
  }
  \label{tbl:configurations}
  \begin{ruledtabular}
    \begin{tabular}{rrcccclD{x}{~}{0.4}l}    
      ID &
      \shortstack[r]{$m_p^{(DM)}$ \\ $[MeV]$} & 
      $\frac{m_0^{(DM)}}{m_0^{(BM)} + m_0^{(DM)}}$ &
      $m^{(BM)} / M_\odot$  & 
      $m^{(DM)} / M_\odot$  &
      $\frac{m^{(BM)} + m^{(DM)}}{ M_\odot}$ &
      \shortstack[r]{$R^{(BM)} / M_\odot$ \\ $(R^{(BM)} [km])$ } &
      \multicolumn{1}{l}{\shortstack[c]{$R^{(DM)} / M_\odot$ \\ $(R^{(DM)} [km])$}} &  
      DM structure
      \\
      \hline
      \ID{1}{id:1_single_fluid} &  $-$  & $0.0 \%$   & 1.27300 & 0.0                              & 1.27300 & 6.4 (9.5) & -                      & $-$     \\
      \ID{2}{id:2_dark_halo}    &  170  & $0.5 \%$   & 1.26641 & $6.75\phantom{0} \times 10^{-3}$ & 1.27316 & 6.4 (9.5) & 11.1 x (         16.4) & halo    \\
      \ID{3}{id:3}              &  250  & $0.5 \%$   & 1.26590 & $6.81\phantom{0} \times 10^{-3}$ & 1.27271 & 6.4 (9.5) & 6.1 x (\phantom{0}9.0) & core    \\
      \ID{4}{id:4}              &  350  & $0.5 \%$   & 1.26557 & $6.86\phantom{0} \times 10^{-3}$ & 1.27243 & 6.4 (9.5) & 4.5 x (\phantom{0}6.6) & core    \\              
      \ID{5}{id:5}              &  500  & $0.5 \%$   & 1.26534 & $6.91\phantom{0} \times 10^{-3}$ & 1.27225 & 6.4 (9.5) & 3.4 x (\phantom{0}5.0) & core    \\              
      \ID{6}{id:6}              &  750  & $0.5 \%$   & 1.26518 & $6.94\phantom{0} \times 10^{-3}$ & 1.27212 & 6.4 (9.5) & 2.5 x (\phantom{0}3.7) & core    \\              
      \ID{7}{id:7}              & 1000  & $0.5 \%$   & 1.26511 & $6.96\phantom{0} \times 10^{-3}$ & 1.27207 & 6.4 (9.5) & 2.0 x (\phantom{0}3.0) & core    \\
      \ID{8}{id:8}              &  350  & $5.0 \%$   & 1.20503 & $6.768 \times 10^{-2}$           & 1.27271 & 6.2 (9.2) & 8.2 x (          12.1) & halo    \\
      \ID{9}{id:9}              &  500  & $5.0 \%$   & 1.20056 & $6.827 \times 10^{-2}$           & 1.26883 &6.2 (9.2) & 5.2 x (\phantom{0}7.7) & core    \\
      \ID{10}{id:10}            &  750  & $5.0 \%$   & 1.19713 & $6.877 \times 10^{-2}$           & 1.26590 &6.1 (9.0) & 3.5 x (\phantom{0}5.2) & core    \\
      \ID{11}{id:11_dark_core}  & 1000  & $5.0 \%$ & 1.19552 & $6.898 \times 10^{-2}$             & 1.26450 &6.1 (9.0) & 2.7 x (\phantom{0}4.0) & core    \\
    \end{tabular}
  \end{ruledtabular}
\end{table*}

{
We consider different configurations by varying DM particle mass, mass fraction of DM and separation between NSs. 
In all configurations the individual NSs have the same total rest mass,
\ie{}, the combined rest mass of BM and DM is $1.4 M_\odot$. In all setups,
the NSs have equal masses and are irrotational, $w^i=0$, \ie{}, 
they are non-spinning. The assumption of vanishing spin is 
reasonable, because NSs spin down, \eg{} due to magnetic
breaking, and the NSs in binary mergers are usually very old,
\ie{} they have spun down for a long time.

We select six values of the DM particle mass in the range
between 170 MeV and 1000 MeV, i.e., 170, 250, 350, 550, 750, and 1000 MeV.
Furthermore we consider configurations with a DM rest mass fraction
of 0\%, 0.5\% and 5\%. In Table~\ref{tbl:configurations} we give
an overview of the different configurations and report the 
properties a corresponding isolated NSs would have.
There we also show the gravitational masses defined as
\begin{equation}
  m^{(s)} := \int_0^{R^{(s)}} 4 \pi r^2 e^{(s)} dr \, ,
\end{equation}
with $R^{(s)}$ the radius of the surface of fluid.
In a binary system the gravitational mass of an individual NS 
can only be defined in a meaningful way in the limit of infinite 
separation, in which the binary components can be viewed as isolated.
Hence we chose to work with fixed baryonic rest masses $m_0^{(s)}$
instead, which is invariantly defined even in binary systems.

The choice of the lowest DM particle mass value, 170 MeV, is motivated by 
the results of Ref.~\citep{PhysRevD.102.063028}, where it 
was shown that for the DM particle masses below 174~MeV DM admixed 
NSs agree with astrophysical observations of the heaviest NSs 
for an arbitrary relative fraction of DM. 
Note, that this is not the case for the higher particle mass, 
where the fraction of DM is constrained in some interval 
(for more details see Ref.~\citep{PhysRevD.102.063028}). 
Moreover, the chosen mass of 
170~MeV and the fraction of 0.5\% leads to a relatively small halo of
approximately twice the radius of the BM component, which is easy to 
model.  When the size of the halos overlap, it is no longer possible 
to fit the element surfaces to the outer fluid of a star.  
Hence, we are discarding the configurations of DM particles with 
DM particle masses of 170~MeV and 250~MeV in the 5.0\% DM case.
Fermionic DM particles with a mass of 1000~MeV present an interesting case,
that resembles nucleons.

We focus on three particular configurations on the extreme opposite
of our parameter spectrum.  Configuration~\ref{id:2_dark_halo} has the
smallest DM particle mass, $m_p^{(DM)} = 170~\mathrm{MeV}$,
and the smallest non-vanishing DM fraction, 0.5\%.  In this configuration
the DM extends beyond the surface of the baryonic fluid and
in figures we consequently label it as the \emph{dark halo} configuration.
On the other side of the spectrum we find 
configuration~\ref{id:11_dark_core} with 
the largest DM particle mass, $m_p^{(DM)} = 1000~\mathrm{MeV}$, 
and a DM fraction of 5\%, for which the DM is concentrated in the core
of the stars. Consequently we label the latter as the 
\emph{dark core} configuration.  We note however, that the name
\emph{dark halo} does not indicate that DM exists only in the surroundings
of the star.  In fact most of the DM is still concentrated in the center
as can be appreciated from Fig.~\ref{fig:SpecificEnthalpy2D}. In the same
way the core of the \emph{dark core} configuration includes a mixture
of BM and DM. The third configuration is the special case of a purely baryonic
star, the \emph{single fluid} configuration (ID~\ref{id:1_single_fluid}),
which we use as a reference.}

We describe BM by a piecewise-polytropic
fit~\citep{Read:2008iy} to the SLy EoS~\citep{Douchin:2001sv}.
As a model of DM, we investigate the degenerate,
relativistic Fermi gas of spin-$\frac{1}{2}$ particles at zero temperature,
for which the EoS is read in as tabulated data.
EoSs at zero temperature are sufficient for our calculations,
because the Fermi energy of the system is
much higher than its temperature.
The typical temperature $T_0$ of NS cores is of the order of
$10^6-10^8$ K~\cite{CumBroFat17,PagGepWeb06}.
We assume that DM has the same temperature as the BM,
because the captured DM particles
keep scattering with baryons, rarely but often enough to
thermalise with the BM component.
A core temperature of approximately $10^8$ K
is much lower than the Fermi energy of BM. This is also true for the Fermi gas
EoS we consider, \eg{} in the \emph{dark halo} case the Fermi energy of DM
reaches 403 MeV in the center of the star,
an energy smaller than that of the BM, but still much larger than the
temperature of the star, $k_B T_0 \approx 0.009$~MeV.
{
In evolutions the neutron stars heat up when they collide, so that
it would become necessary to use finite temperature EoSs. 
This can be achieved by employing EoS tabulated at finite temperatures,
\eg{} the finite-temperature SLy EoS of~\cite{GulRad15, RadGul19},
or by adding a temperature dependent term to the
pressure~\cite{BauJanOec10}.
}

\subsection{Convergence}
\label{Subsection:Convergence}

To validate the code, we check the convergence of the Hamiltonian
constraint for a \emph{dark halo} configuration
of NSs with a separation of 44~$M_\odot$ (65.0~km) on a quasi-circular orbit.

\begin{figure}
  \includegraphics[width=\columnwidth]
  {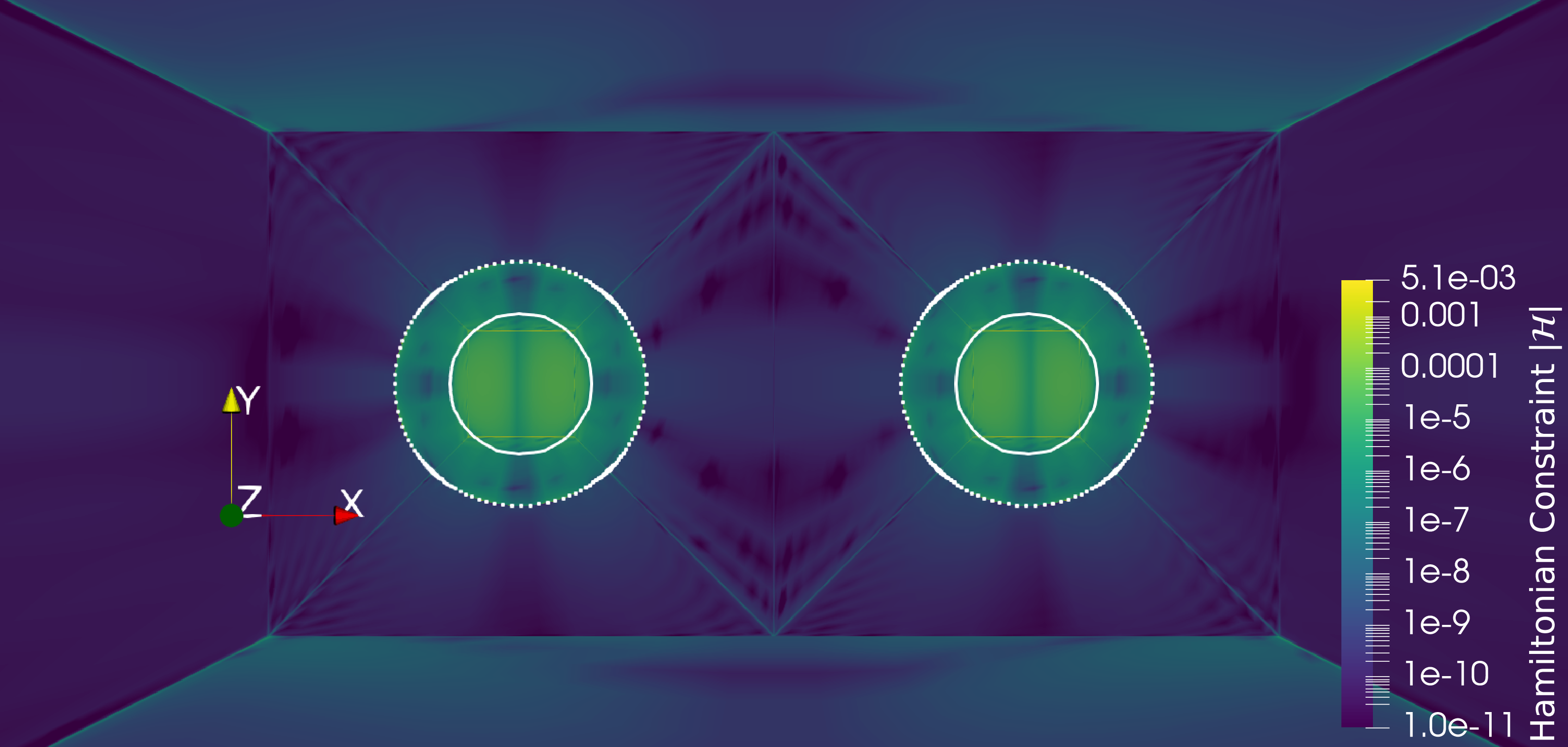}
  \caption{
    \label{fig:HamiltonianConstraint2D} Hamiltonian constraint
    in a dark halo configuration (ID~\ref{id:2_dark_halo}) 
    in the $z=0$ plane.
    White solid outline: surface of the BM fluid.
    White dotted outline: surface of the DM fluid.
  }
\end{figure}

Fig.~\ref{fig:HamiltonianConstraint2D} shows the magnitude of the
Hamiltonian constraint $\mathcal{H}$ on the $z=0$ plane.
The constraint violations are largest in the interior of the star, where
they reach values up to $4 \times 10^{-5}$, whereas in the vacuum regions
the error drops to values below $10^{-9}$, but with some spikes on the order
of $10^{-7}$ at the element boundaries, {which is a behaviour commonly
seen for spectral codes, an example being Fig.~10 of~\cite{TicRasDie19}.
Such spikes in the Hamiltonian constraint usually do not cause 
any problems in subsequent evolutions.
Furthermore the magnitude of these spikes converges towards
zero with increasing resolution.}

The Hamiltonian constraint is largest in the region where the inner
fluid is non-vanishing. In Fig.~\ref{fig:HamiltonianConstraint2D} one can
observe a clear transition on the surface of the baryonic fluid
to lower constraint violations in the DM halo.

\begin{figure}
  \includegraphics[width=\columnwidth]
  {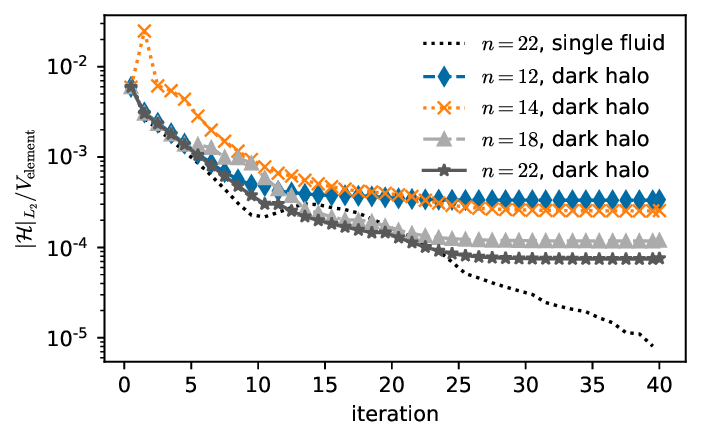}
  \caption{
    \label{fig:HamiltonianConstraintConvergence}
    $L_2$-norm over the inner cube in one of the stars,
    normalised by the volume of the inner cube for {
    \emph{single fluid} (ID~\ref{id:1_single_fluid}) and 
    \emph{dark halo} configurations (ID~\ref{id:2_dark_halo}).}
    The different lines show configurations with different number of points $n$
    in each dimension.     
  }
\end{figure}

Fig.~\ref{fig:HamiltonianConstraintConvergence} demonstrates the development
of the volume-normalised $L_2$-norm of the Hamiltonian constraint
for the inner cube
of one of the stars during the iterative solving process.
The figure shows the behaviour for different number of points $n$ in each
dimension, which is the same for each spectral element.
All curves show a saturation in the norm of the Hamiltonian constraint
towards the end of the iteration process, which for all configurations
is stopped after 40 iterations. Furthermore, it is visible that higher
resolution leads to smaller violations of the Hamiltonian constraint in
the final solution.
For comparison Fig.~\ref{fig:HamiltonianConstraintConvergence} also shows the
sequence
for a corresponding \emph{single fluid} configuration with the same mass and
separation. After 40 iterations {the \emph{single fluid} configuration 
has a Hamiltonian constraint }
10 smaller than the \emph{dark halo} configurations and it does not show
any signs of saturation, \ie{} it would probably reach even smaller constraint
violations if iterated further.
{The reason for this discrepancy is the position of the boundary
of the BM, which in the \emph{dark halo} case lies in the
interior of the spectral element instead of the element surface and 
therefore due to Gibbs' phenomenon requires more resolution to reach 
the same constraint violations.
To improve the efficiency of the method it would be possible to 
introduce an advanced domain decomposition with surface adapted
coordinates for the inner and outer fluid.}

\begin{figure}
  \includegraphics[width=\columnwidth]
  {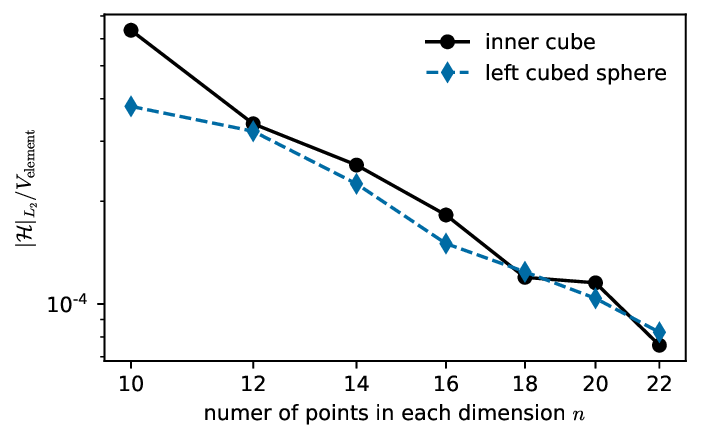}
  \caption{
    \label{fig:HamiltonianConstraintConvergence2}
    Normalised $L_2$-norm of the Hamiltonian constraint in a \emph{dark halo}
    configuration (ID~\ref{id:2_dark_halo}) for a different
    number of points per dimension.
    The norm is normalised by the volume of the spectral element.
    Note that the $x$-axis and $y$-axis are scaled logarithmically.
  }
\end{figure}

The convergence in the final solution is further investigated in
Fig.~\ref{fig:HamiltonianConstraintConvergence2}, which shows its
$L_2$-norm of the Hamiltonian constraint with respect to the number of
collocation points in the spectral elements. The figure shows the
constraint violation for the inner cube element and for the cubed sphere
facing towards the companion star, which is also representative
for all other cubed sphere elements inside the NSs.
The curves are almost straight lines on the log-log-plot of
Fig.~\ref{fig:HamiltonianConstraintConvergence2}, which is compatible with
a polynomial convergence of the constraints, \ie{},
$|\mathcal{H}|_{L_2} \sim n^{-p}$, with $p$ the order of convergence.
This is the expected convergence behaviour for non-smooth data,
which we have due to the surface of the inner fluid.
Using the highest and lowest resolution we can estimate the order of
convergence in the inner cube element to be
$p \approx \log_{22/10} (|\mathcal{H}|_{L_2,n=10} / |\mathcal{H}|_{L_2,n=22} )  \approx 2.7$.

\begin{figure}
  \includegraphics[width=\columnwidth]
  {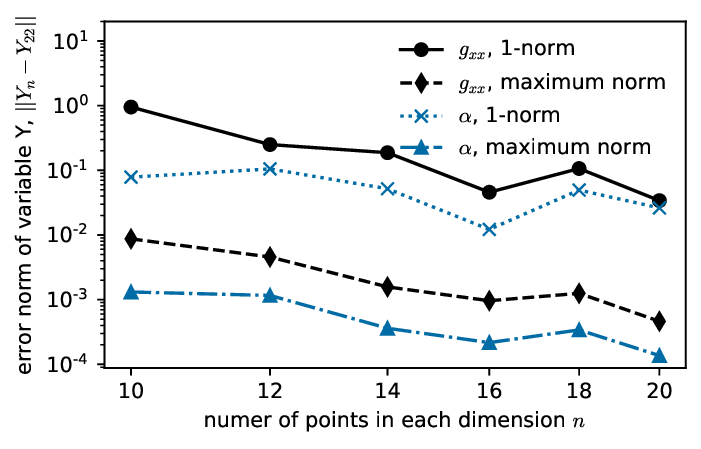}
  \caption{
    \label{fig:MetricConvergence}
    Self-convergence of metric variables in dark halo configurations
    (ID~\ref{id:2_dark_halo}) .
    Black: error norm of the $g_{xx}$ component of the metric.
    Blue: error norm of the lapse, $\alpha$.
    We not that the 1-norm is not normalised by the number of points.
  }
\end{figure}

\begin{figure}
  \includegraphics[width=\columnwidth]
  {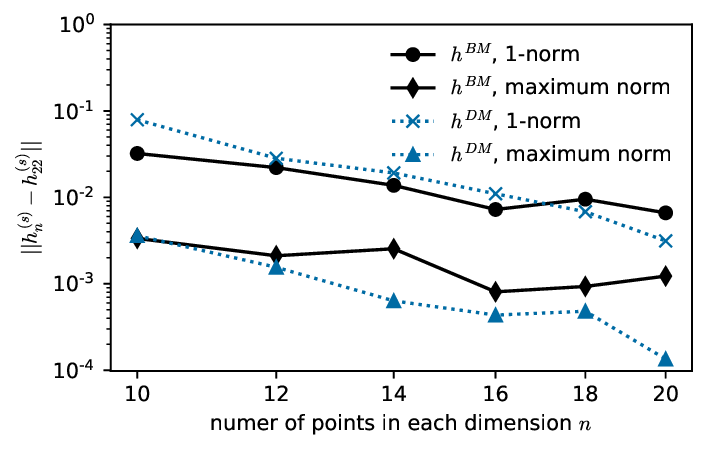}
  \caption{
    \label{fig:hConvergence}
    Self-convergence of the specific enthalpy in dark halo configurations
    (ID~\ref{id:2_dark_halo}).
    Black:
    error norm of the baryonic specific enthalpy $h^{(BM)}$, which is the
    inner fluid.
    Blue: error norm of the specific enthalpy of DM, $h^{(DM)}$.
    We not that the 1-norm is not normalised by the number of points.
  }
\end{figure}

To investigate the convergence of the actual solution variables
we interpolate the data
from different resolutions on a common set of points and compute norms of the
estimated errors on these points.
We interpolate the solution onto a $10 \times 10 \times 10$-grid
equidistant in each direction, with coordinate components given by
$r^i \in \lbrace 20 m/9, m \in [0..9]\rbrace$. This grid
includes some points with pure vacuum, points with only one fluid present
and points with both fluids present.
The error in the solution is estimated by taking the difference to the
solution with the highest resolution, \ie{}, the solution that has 22 points
in each dimension of the spectral elements.
In Fig.~\ref{fig:MetricConvergence} we show the convergence of the
1-norm and the
maximum norm over the set of interpolated points for the $g_{xx}$ component
of the metric and the lapse $\alpha$. Both quantities do not show a monotonic
decay of the error, but there is an overall trend of decaying error.
This somewhat broken convergence behaviour can again be attributed to the
presence of non-smooth fields on the surface of the inner fluid.
Fig.~\ref{fig:hConvergence} shows the convergence of the error
in the specific enthalpy. The DM in this configuration is fitted to the
element boundaries and its specific enthalpy displays a
relatively clear convergence behaviour. The BM fluid on the other hand
shows a very broken convergence and only very little improvement from
the lowest to the highest number of points.
The maximum norm of the error is actually growing
for the two largest number of points, whereas the 1-norm
of the error is also slightly broken, but with an overall behaviour
similar to that of $g_{xx}$ and $\alpha$.

It should be noted, that it is not clear whether the {formalisms used
to construct NS binary initial data actually possesses 
a unique solution and likewise this is true for our formalism
in Sec.~\ref{Section:Formalism}.}
The partial differential equation~\eqref{eq:PDEContinuityEquation}
is not strictly elliptic on the
fluid surface and hence the standard theorems for the uniqueness of the
solution can not be applied. Instead our algorithm might find
\emph{a} solution of many possible, which is another possible
explanation for the slightly broken convergence behaviour.

\subsection{Difference in the Fluid Velocities}
\label{Subsection:Velocities}

It is worth pointing out that even if the BM and DM fluid
components are both irrotational, \ie{}, non-spinning,
the exact velocity profiles are not the same.
The reason for this does not lie in the notion of an irrotational fluid,
but is caused by differences in the fluids' equations of motion.
An irrotational fluid~\cite{Shi98,Tic11,BauSha10} is defined by the vanishing of its kinematic vorticity
tensor~\cite{HawEll73a}
\begin{equation}
  \omega_{\alpha \beta} := P^\mu_\alpha P^\nu_\beta \nabla_{[\mu} u_{\nu]} = 0\, ,
\end{equation}
with $P^\mu_\alpha = \delta^\mu_\alpha + u^\mu u_\alpha$ {and its 
rotational component, $w^{(s)i}$ in Eq.~\ref{eq:VelocitySplit}, vanishes.}
This notion
does not depend on the thermodynamic properties of the fluid
and hence differences in the velocities can only be the result of the
of the equations of motion used in the derivation of the formlalism
in Sec.~\ref{Section:Formalism}, \ie{} the Euler equations~\cite{Shi98,Tic11,BauSha10}
\begin{equation}
  u^{(s)\mu} \nabla_\mu (h^{(s)} u^{(s)}_\nu + \nabla_\nu h^{(s)}) = 0 \, ,
\end{equation}
which follow from $\nabla^\mu T^{(s)}_{\mu\nu} = 0$, and the continuity equation
\begin{equation}
  \nabla_\mu (\rho^{(s)}_0 u^{(s)\mu}) = 0 \, .
\end{equation}
If for example
the DM would have the same four-velocity as the BM, it would still be
irrotational, but might be incompatible with the laws of energy-momentum
or particle number conservation.

In nature the disparity in the fluid velocities is affected by two
counter-acting effects, particle scattering between BM and DM on the one hand
and physics determining spin-down on the other hand.
In our formulation the two fluids are modelled as non-interacting, but the
BM-DM scattering cross-section might be non-zero in nature,
which would drive the two fluids towards a common velocity.
This process is counter-acted by effects driving the fluid into an irrotational
state, as for example magnetic braking for BM~\cite{OstGun69,ColPosPop01,
RogSaf16}.
It is unclear whether a
similar effect exists for DM and whether it is dominant over the effect of
BM-DM scattering. By assuming vanishing of the kinematic vorticity
for the DM component, we assume that such an effect exists and it is also
dominating over the scattering with BM.

\begin{figure}
  \includegraphics[width=\columnwidth]
  {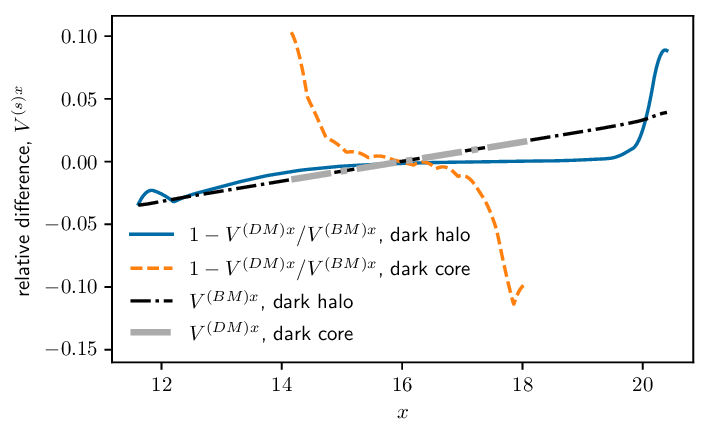}
  \caption{
    \label{fig:Velocities} Relative difference in the velocities
    for {configuration~\ref{id:2_dark_halo} (\emph{dark halo})
    and~\ref{id:11_dark_core} (\emph{dark core})  }
    with a separation of 32~$M_\odot$.
    The difference is shown along a diagonal with the parametrization
    $r^i(s) = s (1,1,0) + r_c^i$, going
    through the center of the star located at $r_c^i = (16 M_\odot,0,0)$.
    $V^{(BM)x}$ (black, {dash-dotted} line) and $V^{(DM)x}$ 
    (grey, dash-dotted line)
    show the $x$-component of the velocity of the respective inner fluid.
  }
\end{figure}

We find that both fluids move with basically the same velocity,
with coinciding velocities in the star center, but increasing
difference towards the surface of the inner fluid.
We quantify this effect in terms of the residual three-velocity $V^{(s)i}$,
in which the orbital movement given by the Killing vector $\xi^\mu$
is split off,
\begin{equation}
  V^{(s)i} = u^{(s)i}/u^{(s)0} - \xi^i \, .
\end{equation}
Fig.~\ref{fig:Velocities} shows the $x$-component of $V^{(s)i}$
and the relative difference
of the fluid velocities for the region in which both fluids are present.
We present results for configurations at a separation of 32~$M_\odot$, 
a separation at which the DM halos in the \emph{dark halo} configurations
(ID~\ref{id:2_dark_halo})
are already relatively close and deformed, {as we demonstrate in
Fig.~\ref{fig:SpecificEnthalpy2D}.}
We find that differences in the two fluids are smaller for larger separation,
which is intuitively understandable, because for large separations the
system goes to the limit of isolated NSs in which the fluid velocities
coincide. 

The data in Fig.~\ref{fig:Velocities} is shown along a diagonal
through the star parametrized in the following way:
$r^i(s) = s (1,1,0) + r_c^i$, where $r_c^i$ is the center of the star.
We choose to present the data along this diagonal because the difference
$V^{(BM)i} - V^{(DM)i}$ has a quadrupolar structure with nodes
going through $r_c^i$ and being approximately parallel to 
the $x$- and $y$-axes.
Hence the difference is basically zero on the $x$- and $y$-axes,
but very prominent along the specified diagonal.
The relative difference between the residual velocities is below 0.2\%
near the center of the star and reaches up to 10\% on the surfaces of the
inner fluids.
The difference between the velocities of the \emph{dark halo}
{(ID~\ref{id:2_dark_halo})}
and \emph{dark core} {(ID~\ref{id:11_dark_core})} configurations is 
relatively small,
which can be seen from the fact the curves of the velocities of the 
inner fluids lie on top of each other.

\subsection{Binding Energy}
\label{Subsection:BindingEnergy}

NSs with a DM component are more tightly bound,
because the DM component adds gravitating mass,
but provides no additional repulsion to balance the gravitational
pressure. {This effect is well studied and was already demonstrated by several 
authors~\cite{PhysRevD.102.063028,2011PhRvD..84j7301L,Xiang:2013xwa}.
In the following we investigate the effect of DM on the
energetics of the binary system, \ie{} the orbital binding energy.}

\begin{figure}
  \includegraphics[width=\columnwidth]
  {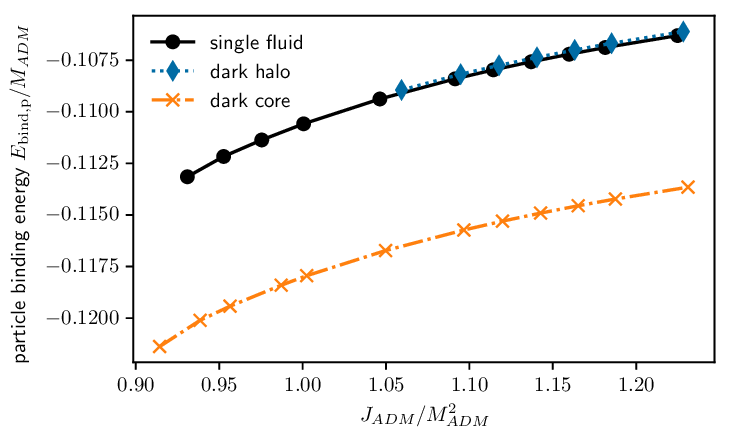}
  \caption{
    \label{fig:BindingEnergy} Particle binding energy $E_\mathrm{bind,p}$
    as a function of the ADM angular momentum. {We show results for
    configuration~\ref{id:1_single_fluid} (\emph{single fluid}),
    \ref{id:2_dark_halo} (\emph{dark halo})
    and~\ref{id:11_dark_core} (\emph{dark core}).}
  }
\end{figure}

The gravitational binding energy of the particles
is the difference of the ADM mass~\cite{ArnDesMis62,Gou12,BauSha10}
and the sum of the rest masses $m_{0i}^{(s)}$ of the components.
If all fluid particles would fall in from infinity, the true ADM mass would
equal the total rest mass.
However, the configurations that we construct do not contain
GWs and therefore they do not model the energy lost in
gravitational radiation. The difference in our ADM mass estimate and the
total rest mass is, therefore, a measure of the particle binding energy:
\begin{equation}
 E_\mathrm{bind,p} = M_{ADM}
   - m_{01}^{(BM)}-m_{01}^{(DM)} - m_{02}^{(BM)}-m_{02}^{(DM)} \, .
\end{equation}
{We have constructed the configurations with fixed baryonic masses,
but configurations with different separation distance between the stars or particle mass $m_p^{(DM)}$
will have a different ADM mass, similar to how the isolated stars
in Table~\ref{tbl:configurations} have different gravitational masses.
To make the results comparable in the
figures we show quantities appropriately rescaled by $M_{ADM}$.

We construct a series of configurations with varying orbital separation $D$.
The orbital frequency $\Omega$ changes as well, since it is a function 
of the masses and the orbital separation.}
Fig.~\ref{fig:BindingEnergy} shows the {rescaled}
particle binding energy as a function
of our estimate for the ADM angular momentum $J_{ADM}$.
It can be seen that \emph{dark core} configurations 
{(ID~\ref{id:11_dark_core})} are
more tightly bound than \emph{single fluid} configurations. The
\emph{dark halo} configurations {(ID~\ref{id:2_dark_halo})}
seemingly coincide with the single fluid case.
This can be attributed to the relatively low DM fraction of only 0.5\% in these
configurations.
All configurations are more tightly bound for smaller $J_{ADM}$ corresponding
to smaller stellar separations. This is due to the stronger
orbital binding between the two stars.

\begin{figure}
  \includegraphics[width=\columnwidth]
  {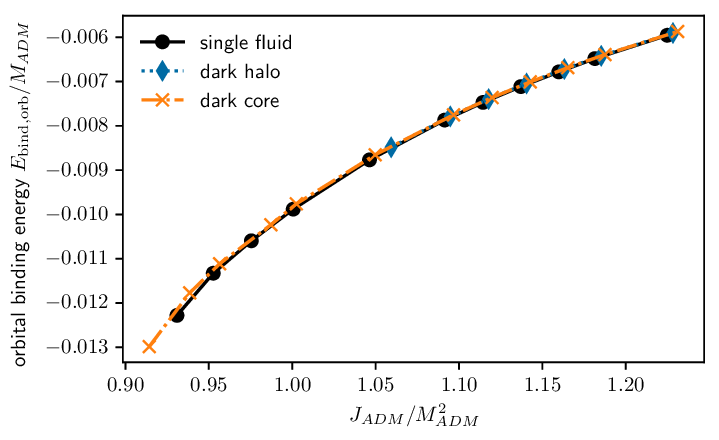}
  \caption{
    \label{fig:OrbitalBindingEnergy} Orbital binding energy
    $E_\mathrm{bind,orb}$ as a function of the ADM angular momentum.
    {We show results for
    configuration~\ref{id:1_single_fluid} (\emph{single fluid}),
    \ref{id:2_dark_halo} (\emph{dark halo})
    and~\ref{id:11_dark_core} (\emph{dark core}).}
  }
\end{figure}

\begin{figure}
  \includegraphics[width=\columnwidth]
  {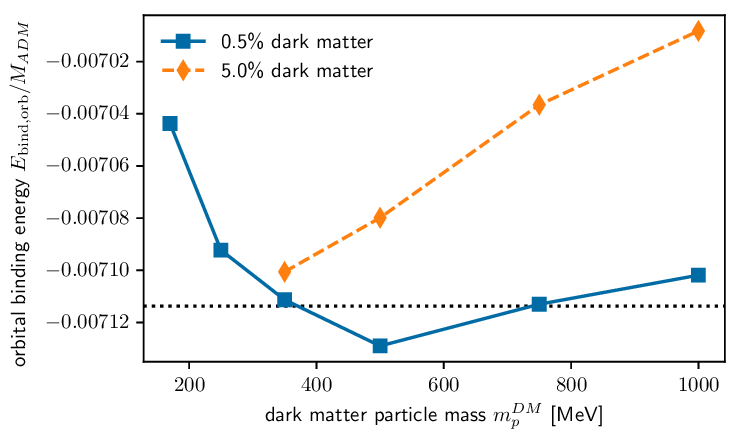}
  \caption{
    \label{fig:Ebind_vs_mD} {Orbital binding energy $E_\mathrm{bind,orb}$ as a function of the DM particle mass 
    $m_\mathrm{DM}$ for a binary separation of 36~$M_\odot$ (53.2~km). 
    As a reference the horizontal
    black dotted line shows the value for the \emph{single fluid} %
    configuration (ID~\ref{id:1_single_fluid}).}
  }
\end{figure}

\begin{figure}
  \includegraphics[width=\columnwidth]
  {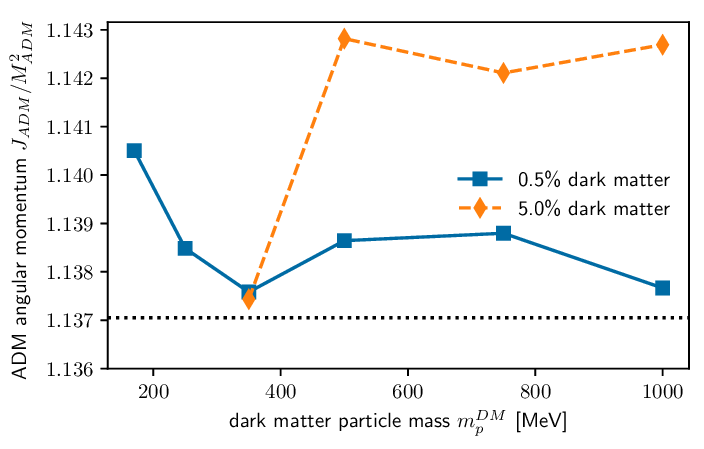}
  \caption{
    \label{fig:J_vs_mDM} {Angular momentum as a function of the DM particle mass 
    $m_\mathrm{DM}$ for a binary separation of 36~$M_\odot$ (53.2~km). As a reference the horizontal
    black dotted line shows the value for the \emph{single fluid} %
    configuration (ID~\ref{id:1_single_fluid}).}
  }
\end{figure}

Most of the binding energy is contained in the individual stars
and the contribution of the orbital binding energy is universal in all
configurations.
The orbital binding energy $E_\mathrm{bind,orb}$ is the energy necessary
for the two NSs to escape to infinity. It can be computed using
the gravitational mass $m_{i}^{(s)}$ of the components, by
\begin{equation}
 E_\mathrm{bind,orb} = M_{ADM}
    - m_{1}^{(BM)}-m_{1}^{(DM)} - m_{2}^{(BM)}-m_{2}^{(DM)} \, .
\end{equation}
The gravitational masses $m_{i}^{(s)}$ are obtained by solving a
TOV-like equation for isolated stars that have the same rest masses.
The gravitational mass $m_{i}^{(s)}$ is smaller than
the rest mass $m_{i0}^{(s)}$, because it accounts for the binding energy.
Hence, $E_\mathrm{bind,orb}$ contains only contributions of the
binding energy that are due to the mutual binding between the stars.
Fig.~\ref{fig:OrbitalBindingEnergy} shows that {the relation between
orbital binding energy and ADM angular momentum
(both appropriately rescaled by $M_{ADM}$) 
is mostly independent of the DM configuration as the lines are falling on
top of each other.

To investigate the small effect of the particle mass, we construct
configurations at a range of DM particle masses $m_p^{(DM)}$ 
from 170 MeV to 1000 MeV 
at a fixed binary separation of 36~$M_\odot$ (53.2~km)
Fig.~\ref{fig:Ebind_vs_mD} shows the orbital binding energy for
the case of 0.5\% and 5\% of DM. For the case of 0.5\% DM 
(IDs~\ref{id:2_dark_halo} to~\ref{id:7})
we find that the orbital binding energy has a minimum around 550 MeV.
The value of the minimum lies even below that of the corresponding
\emph{single fluid} configuration.
For the case of 5\% of DM (IDs~\ref{id:8} to~\ref{id:11_dark_core}) 
we do not observe a minimum, but an orbital
energy always larger than in the corresponding
\emph{single fluid} configuration and
increase roughly linear with $m_p^{(DM)}$.
We emphasise that for this comparison one has to keep in mind that
configurations with different $m_p^{(DM)}$ also have different 
angular momentum. However, as is shown in Fig.~\ref{fig:J_vs_mDM}
the variation in the rescaled ADM angular momentum is below 1\%.
We also find no clear relation between the ADM angular momentum
and $m_p^{(DM)}$, but we find that larger amounts of DM
tend to lead to larger angular momentum.

Figures~\ref{fig:Ebind_vs_mD} and~\ref{fig:J_vs_mDM} also demonstrate
that by decreasing the amount of DM the configurations
reach the single fluid limit. In almost all cases the configurations 
with lower DM fraction have a binding energy and angular momentum that
is closer to that of the single fluid case.
Only for the particle masses of 350 MeV 
the angular momentum of the 5.0\% configuration is closer to 
the single fluid limit. However, it must be noted that for this case 
the DM forms a halo around the BM and therefore this configuration is not 
representative for the limit of low DM content.
}

\subsection{Deformation}
\label{Subsection:Deformation}

To quantify the deformation of the stars we compute the ratio of the
diameters along the orbital radius and along the polar axes.
The diameter along the orbital radius is taken as $\Delta x$,
largest difference in the $x$-coordinates of two points on the fluid surface.
The polar diameter $\Delta z$, is the largest difference in the
$z$-coordinate of two points on the fluid surface.
The tidal force of the companion stretches the star in $x$-direction,
whereas the poles are slightly flattened. This measure of deformation
is of course coordinate-dependent, but it still provides some physical insights.

\begin{figure}
  \includegraphics[width=\columnwidth]
  {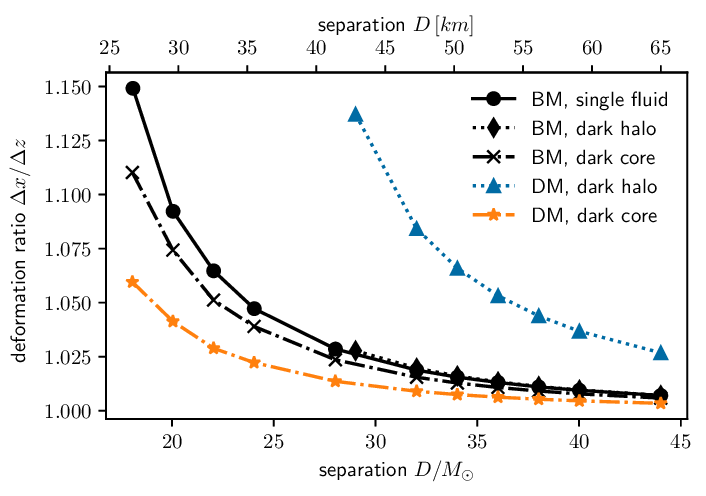}
  \caption{
    \label{fig:chiR} Deformation $\Delta x / \Delta z$ of the fluid surfaces
    as function of the NS centres.
    The deformation is computed as the ratio of the largest extents in
    $x$ and $z$ direction.
    Curves labeled BM show the deformation of the surface of the baryonic
    fluid, whereas curves labeled DM show the deformation of the DM surface.
    {We show results for
    configuration~\ref{id:1_single_fluid} (\emph{single fluid}),
    \ref{id:2_dark_halo} (\emph{dark halo})
    and~\ref{id:11_dark_core} (\emph{dark core}).}
  }
\end{figure}

{We analyse the same set of configurations with varying orbital 
separation $D$ as in the previous section.}
Fig.~\ref{fig:chiR} shows the deformation $\Delta x / \Delta z$
for each fluid surface. When the NSs are closer,
the tidal forces on the companion are stronger and hence the
deformation is stronger. It can be observed that NSs with a DM core are
systematically less deformed than their one-fluid counterparts.

The strong deformation in the \emph{dark halo} 
case (ID~\ref{id:2_dark_halo}) can also be seen in
Fig.~\ref{fig:SpecificEnthalpy2D}, which shows a cut through the $z=0$ plane.
For a separation of 32~$M_\odot$ (47.3~km) the deformation is clearly
visible by eye.
At a separation of 28~$M_\odot$ (41.3~km) the deformation becomes already so
strong that the surfaces of the NSs touch and mass shedding occurs.

\begin{figure}
  \includegraphics[width=\columnwidth]
  {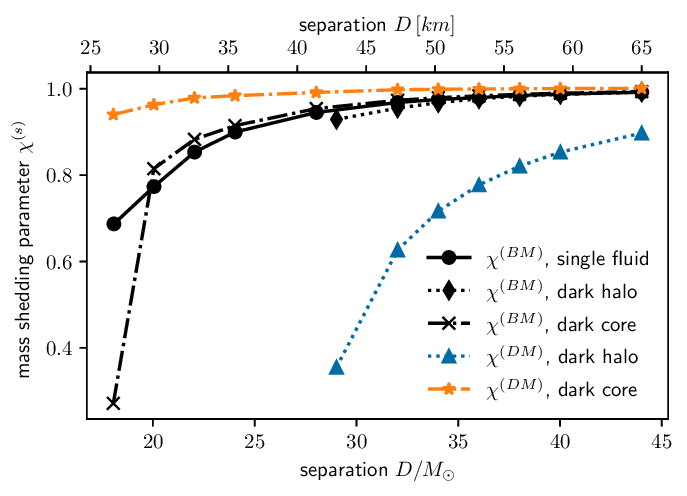}
  \caption{
    \label{fig:chi} Mass shedding parameter $\chi$
    as a function of the separation of the NS.
    {$\chi^{(BM)}$ is computed from} the deformation of the surface of the baryonic
    fluid, whereas {$\chi^{(DM)}$ is computed from} the DM surface.
    {We show results for
    configuration~\ref{id:1_single_fluid} (\emph{single fluid}),
    \ref{id:2_dark_halo} (\emph{dark halo})
    and~\ref{id:11_dark_core} (\emph{dark core}).}
  }
\end{figure}

The closeness to mass shedding can be quantified
in terms of the mass-shedding parameter $\chi$, which was
first introduced in~\cite{GouGraTan01} and which we define as
\begin{equation}
  \label{eq:chi}
  \chi^{(s)} = \frac{\partial_x h^{(s)}|_\mathrm{eq}}{\partial_z h^{(s)}|_\mathrm{pole,avg}} \, ,
\end{equation}
where the label ``eq'' denotes the point on the surface, which is
facing towards the companion star and for which the $x$-coordinate
is extremal. The label ``pole'' denotes the surface points at which the
$z$-coordinate is extremal and where in Eq.~\eqref{eq:chi} the label
``avg'' indicates that we have averaged over the values at the
``north and south pole''. Note that for non-spinning stars the
``north'' and ``south pole'' values only differ slightly due to round-off error.
In the mass shedding limit $\chi^{(s)}$ will tend to 0.
We evaluate the $\chi^{(s)}$ for each fluid component
individually on the respective fluid surfaces.
We show the resulting $\chi^{(s)}$ as a function of the distance
of the centres of the stars in Fig.~\ref{fig:chi}.
The DM fluid in the \emph{dark halo} scenario {(ID~\ref{id:2_dark_halo})}
is easily deformable,
which leads to a relatively small mass shedding parameter of 0.9 already at
a separation of 44~$M_\odot$. We find that a separation of 28~$M_\odot$ leads
to a configuration with touching star surfaces, from which we conclude that
mass shedding occurs somewhere at a separation between 28 and 29~$M_\odot$,
which means the system will
transition relatively slowly to the mass shedding regime over a time
where the two NSs decrease their separation by 16~$M_\odot$.
For the \emph{dark core} configurations {(ID~\ref{id:11_dark_core})},
on the other hand, the transition
to mass shedding is rather sudden with $\chi$ reaching a value of 0.9 at
separation of approximately 23~$M_\odot$ and the mass shedding occurring
for the baryonic fluid at a separation of 16~$M_\odot$.

\begin{figure}
  \includegraphics[width=\columnwidth]
  {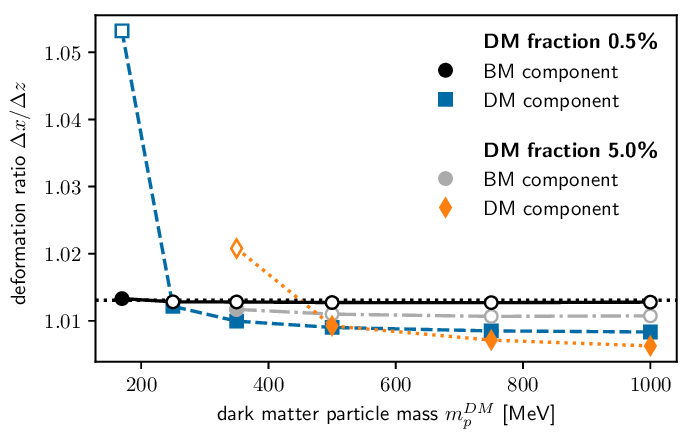}
  \caption{
    \label{fig:deformation_chiR_vs_mD} 
    {Deformation $\Delta x / \Delta z$ as a function of the DM
    particle mass 
    $m_\mathrm{DM}$ for a binary separation of 36~$M_\odot$ (53.2~km). 
    Open symbols denote the fluids with the larger diameter, \ie{} the
    halo component. Filled symbols denote the inner component, 
    \ie{} the core.
    As a reference the horizontal
    black dotted line shows the value for the \emph{single fluid} %
    configuration (ID~\ref{id:1_single_fluid}).}
  }
\end{figure}

\begin{figure}
  \includegraphics[width=\columnwidth]
  {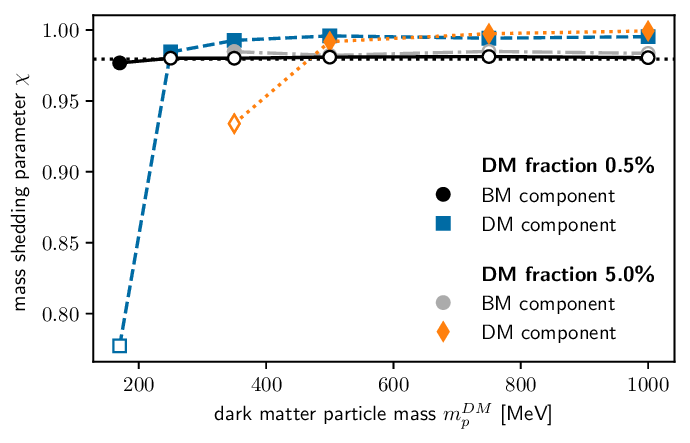}
  \caption{
    \label{fig:deformation_chi_vs_mD} 
    {Mass shedding parameter $\chi$ as a function of the DM 
    particle mass 
    $m_\mathrm{DM}$ for a binary separation of 36~$M_\odot$ (53.2~km). 
Open symbols denote the fluids with the larger diameter, \ie{} the
    halo component. Filled symbols denote the inner component, 
    \ie{} the core.    
    As a reference the horizontal
    black dotted line shows the value for the \emph{single fluid} %
    configuration (ID~\ref{id:1_single_fluid}).}
  }
\end{figure}

{In Figs.~\ref{fig:deformation_chiR_vs_mD} 
and~\ref{fig:deformation_chi_vs_mD} we show the 
deformation~$\Delta x / \Delta z$ and mass shedding 
parameters~$\chi^{(s)}$ as functions of the DM particle mass~$m_p^{(DM)}$
corresponding to all configurations in Table~\ref{tbl:configurations}
and for a fixed binary separation of 36~$M_\odot$ (53.2~km).
For the case of 0.5\% DM the BM deformation as well as $\chi^{(BM)}$ 
have practically the same value as in the \emph{single fluid} case.
For 5\% of DM the BM deformation is smaller and $\chi^{(BM)}$ is larger
owing to the higher compactness in these configurations.
The DM fluid is less strongly deformed when $m_p^{(DM)}$ is large.
In particular when the DM fluid changes character from being the halo to 
being the core component, the deformation decreases rapidly and
$\chi^{(DM)}$ increases rapidly.}

\section{Conclusion}
\label{Section:Conclusion}

We have extended the \texttt{SGRID} code to construct constraint-solved,
quasi-equlibrium configurations of binaries of NSs consisting of two
non-interacting fluids.
The second fluid represents DM that can comprise some part of
the matter of NS. In this study we have used the EoS
of a degenerate, relativistic Fermi gas with different particle masses
to model the DM fluid.
These quasi-equlibrium configurations can be used as initial data
for NR inspiral simulations of DM admixed NS binaries. The \texttt{BAM}
code can already evolve mirror DM~\cite{Emma:2022xjs}
and could be easily extended to allow for general EoS for
the DM fluid.

Another possible application of the two fluid approach are
superfluid NS cores.
At sufficiently high density BM forms a state made of
superfluid neutrons and superconducting protons,
which can be described in a two fluid approach.
However, the two fluids still interact with each other
due to the entrainment effect
and the condition of beta-equilibrium~\cite{ComJoy03}.
{The evolution equations for interacting multifluid systems have been
discussed in~\cite{AndCom21, And21}, but so far no formalism exists for the 
construction of initial data for NS binary systems. 
For the construction of such initial data the formalism in this work
could be extended using an interaction model similar to the one used 
in solutions of isolated NS with superfluid cores~\cite{PriNovCom05,Cha08} and taking into account mutual
friction~\cite{AndSidCom06}.}
In binary NS collisions the temperature will rise above the critical
temperature for superfluidity and superconductivity. {The case of 
finite temperature superfluid dynamics was discussed in~\cite{AndKruCom12}.}

We have tested the convergence of the constructed configurations with respect to
resolution. The Hamiltonian constraint converges polynomially with an order
of $\approx 2.7$. The lack of exponential convergence can be attributed to
the presence of the non-smooth transition of the density at the surface of
the inner fluid, which is not fitted to the boundaries of the spectral elements.
Self-convergence tests for metric components and the specific enthalpies
show that the solution improves with increasing resolution, but with
a slightly broken convergence towards higher resolution, which we again
attribute to the surface of the inner fluid.
For future improvements to the code it is a worthwhile consideration
to implement a new grid layout that allows fitting to the surface of
a second fluid

We have shown that the two fluids do not have the exact same velocities,
but that the difference in the residual velocities reaches up to
10\% on the surface of the inner fluids.
The difference in the velocity profiles will be even stronger if
one assumes independent rotational states for the components.
In this work we only investigated only purely irrotational configurations,
but our formalism, in principle, allows to construct configurations
with arbitrary spin for the individual stars and fluid components.
This is relevant in particular for the DM component, which might
only have insufficient mechanisms to lose angular momentum and hence could
be in a state of rapid rotation.

The presence of DM affects the compactness and deformability
of NSs, which will change the merger dynamics. We have shown
that the presence of DM can delay the point of mass-shedding
to a later stage of the inspiral, \ie{}, towards closer separations.
This is in accordance with the findings in numerical evolutions of
two-fluid binary mergers~\cite{Emma:2022xjs}.
In the case of a DM halo, mass shedding could occur much earlier than
for the baryonic component. However the matter contained in the DM
halo is rather low and hence the impact of DM mass shedding
on the dynamics of the BM is potentially small,
nevertheless, dynamical simulations are needed to verify this assumption.

\begin{acknowledgments}

This work was supported by funding from the FCT – Fundação para a Ciência e a Tecnologia, I.P., within the Project No. EXPL/FIS-AST/0735/2021. H.R.R. and V.S. also acknowledge the support from the project No. UIDB/04564/2020, and UIDP/04564/2020.
W.T. acknowledges funding from the National Science Foundation under grant
PHY-2136036.

\end{acknowledgments}
\bibliography{two_fluid_sgrid.bbl}
\end{document}